\documentclass{ws-ijgmmp}
\usepackage{color}
\begin{document}

\markboth{F. Ahmed}
{Gravitational lensing in Eddington-inspired Born-Infeld gravity background}

%
\catchline{}{}{}{}{}
%

\title{Gravitational lensing in a space-time with cosmic string within the Eddington-inspired Born-Infeld gravity
}

\author{Faizuddin Ahmed\footnote{email: \bf faizuddinahmed15@gmail.com}
}

\address{Department of Physics, University of Science \& Technology Meghalaya, Ri-Bhoi, 793101, India
}

\maketitle

\begin{abstract}
This study explores the deflection angle of photon rays or light-like geodesics within the framework of Eddington-inspired Born-Infeld (EiBI) gravity background space-time, taking into account the influence of cosmic strings. The primary focus lies in deriving the effective potential of the system applicable to both null and time-like geodesics, as well as determining the angle of deflection for light-like geodesics. Our analysis shows that the presence of cosmic strings induces modifications in these physical quantities, leading to shifts in their respective values.
\end{abstract}

\keywords{Modified theories of gravity; Linear defects: dislocations, disclinations; Gravitational lenses and luminous arcs.
}

\section{Introduction}

Gravitational lensing refers to the phenomenon observed when a massive object or a concentration of matter, like a galaxy cluster, creates a gravitational field that produces significant curvature in space-time. Consequently, the path of light originating from a distant source deviates from a straight line and becomes curved as it travels towards the observer. This bending of light is a consequence of the gravitational field's effect on the trajectory of photons, altering the apparent position and characteristics of the source as observed from Earth. Following Eddington's groundbreaking observation of light bending, numerous mathematical investigations have been conducted on gravitational lensing, encompassing both weak and strong limits. These studies have delved into the lensing phenomena occurring in various spacetimes, including those associated with black holes, wormholes, and naked singularities. Through rigorous mathematical analyses, researchers have sought to unravel the intricate mechanisms and effects of gravitational lensing in these exotic space-time configurations.

In the realm of modified gravity, a compelling theory has emerged known as Eddington-inspired Born-Infeld (EiBI) gravity \cite{kk1, kk2}. This theory draws inspiration from both the Eddington gravitational action \cite{kk3} and Born-Infeld nonlinear electrodynamics \cite{kk4}, generating significant interest in the scientific community. EiBI gravity adopts a Palatini-type formulation, where the metric tensor and the connection are treated as independent variables. Unlike conventional approaches, EiBI gravity couples the Eddington action to matter without strictly adhering to a purely affine action. While it reduces to standard General Relativity (GR) in a vacuum, it exhibits distinct gravitational field behavior in the presence of matter. Remarkably, EiBI gravity produced geodesic singularities even within classical treatments while remaining asymptotically compatible with GR. The study of compact star structures within EiBI theory, particularly in nuclear astrophysics, has received considerable attention \cite{kk5, kk6, kk7, kk8, kk9, kk10}. Investigations into the existence of neutron stars \cite{kk2, kk11} have imposed constraints on the Eddington coupling parameter $\epsilon$. Notably, positive values of $\epsilon$ entail an effective gravitational repulsion, leading to the emergence of pressure-less stars composed of non-interacting particles. These pressure-less stars offer intriguing models for self-gravitating dark matter and contribute to expanded mass limits for compact stars \cite{kk11, kk12, kk13}. Recent studies have addressed into various phenomena within EiBI gravity, including the density profile of pressure-less dark matter \cite{kk14}, wormhole geometries \cite{kk15}, the gravitational field of global monopoles \cite{kk16}, and electrically-charged black holes \cite{kk17, kk18}. Additionally, investigations have explored Geonic black holes and remnants \cite{kk20}, non-linear $\sigma$-models \cite{kk21}, quantum effects near classical big rip singularities \cite{kk22}, and Kundt wave geometries \cite{kk23}. For a comprehensive overview of EiBI gravity, readers are encouraged to consult reference \cite{kk19}.

Gravitational lensing is a crucial and significant area of study in cosmology and gravitation. It occurs in both weak field scenarios, where light rays pass far from the source, and strong field situations, where light rays pass very close to massive objects. Extensive investigations have been conducted on gravitational lensing, exploring a diverse range of space-time configurations. These studies include gravitational lensing by charged black holes in the context of string theory \cite{aa1}, brane-world black holes \cite{aa2}, Reissner-Nordstrom black holes \cite{aa4}, naked singularities in space-times \cite{aa6,aa7,aa8,aa9}, Schwarzschild black holes \cite{aa10}, Kerr-Randers optical geometry \cite{aa13}, rotating global monopole space-times \cite{aa14,aa24}, a general asymptotically flat, static, and spherically symmetric space-times \cite{aa15}. Further investigations have explored gravitational lensing in cosmic string space-time under Lorentz symmetry breaking effects \cite{aa19}, with the inclusion of cosmological constant effects in rotating cosmic string space-time \cite{aa22}, Kerr-Newman-Kasuya space-time \cite{aa23}, and global monopoles within the framework of Eddington-inspired Born-Infeld theory \cite{aa32}. Additional studies have focused on gravitational lensing in Kerr-MOG black holes \cite{aa34}, Simpson-Visser black-bounce space-time \cite{aa35}, rotating regular black holes \cite{aa37}, black-bounce space-time \cite{JRN}, stationary axisymmetric space-time \cite{aa38}, and black holes \cite{aa39}. Moreover, gravitational lensing has been examined in topological charged space-time within the framework of Eddington-inspired Born-Infeld gravity \cite{ARS}, spherically symmetric and static space-time \cite{aa26}, Rindler modified Schwarzschild black holes \cite{bb1}, and topologically charged Eddington-inspired Born-Infeld space-time \cite{aa33}.

Furthermore, gravitational lensing has been the subject of extensive investigation in various other geometric backgrounds specifically on wormholes space-time, expanding our understanding of this phenomenon. These studies encompass gravitational lensing in wormholes \cite{aa1, aa3, RS, RS2}, Damour-Solodukhin wormholes \cite{aa36}, rotating wormholes \cite{aa16,aa21}, charged wormholes within the framework of Einstein-Maxwell-dilaton theory \cite{aa17}, Morris-Thorne wormholes \cite{aa25}, traversable Lorentzian wormholes \cite{aa27}, Ellis wormholes \cite{aa18,aa28,aa20,aa40,aa411,aa43}, inclusion of wave effect on gravitational lenses by the Ellis wormhole \cite{aa42}, microlensing by the Ellis wormhole \cite{FB,FB2}, asymptotically conical Morris-Thorne wormholes \cite{aa29,aa31}, and topologically charged Ellis-Bronnikov-type wormholes \cite{aa30}. The lensing properties of symmetric and asymmetric wormholes \cite{aa41}, exponential wormhole spacetimes \cite{bb4}, phantom wormholes using the Gauss-Bonnet theorem \cite{bb2}, and massless wormholes in massive gravity \cite{aa44} have also been studied. These extensive investigations have deepened our understanding of gravitational lensing phenomena in diverse and exotic space-time scenarios.

The line-element describing cosmic string or wormhole space-time in EiBI gravity background in the spherical coordinates $(t, r, \theta, \phi)$ is given by ($c=1=\hbar=G$)
\begin{equation}
ds^2=-dt^2+\frac{dr^2}{\Big(1+\frac{\epsilon}{r^2}\Big)}+r^2\,(d\theta^2+\alpha^2\,\sin^2 \theta\,d\phi^2),
\label{1}
\end{equation}
where $\epsilon$, the parameter associated with the nonlinearity of the EiBI gravity, and $\alpha <1$ is the cosmic string parameter related with linear mass density of the string. The coordinates are in the ranges $0 \leq r < \infty$, $0 < \theta <\pi$, and $0 \leq \phi < 2\,\pi$. One can see from the above line-element that for $\epsilon=0$, the metric becomes a cosmic string space-time in the spherical system given by $ds^2=-dt^2+dr^2+r^2\,(d\theta^2+\alpha^2\,\sin^2 \theta\,d\phi^2)$ Refs. \cite{AV,GAM,GAM2,GAM3}. It is worth mentioning here that the gravitational lensing effects of vacuum strings was investigated in Ref. \cite{JRG}. Similarly, for $\epsilon=-b^2<0$ and $\alpha \to 1$, we have an Ellis-Bronnikov-type wormhole space-time Refs. \cite{HGE,KAB} in which gravitational lensing has widely been investigated Refs.\cite{aa18,aa28,aa20,aa42,aa40,aa411,aa43}. Finally, for $\epsilon=-b^2<0$, one will have Morris-Thorne-type wormhole with cosmic strings which has recently been discussed in Ref. \cite{aa31}. For $\epsilon <0$, one can do comparison of this line-element (\ref{1}) with the general form of Morris-Thorne wormhole metric with cosmic strings \cite{MT,MT2}, that is, $ds^2=-e^{2\,\Phi(r)}\,dt^2+\frac{dr^2}{\Big(1-\frac{A(r)}{r}\Big)}+r^2\,(d\theta^2+\alpha^2\,\sin^2 \theta\,d\phi^2)$ and show that the redshift function is null, $\Phi(r)=0$, and the shape function is given by $A(r)=-\frac{\epsilon}{r}$. This EiBI gravity background space-time with cosmic strings is asymptotically flat since $\frac{A(r)}{r} \to 0$ at $r \to \infty$.

The introduction of a cosmic string into a space-time brings about significant alterations to its geometry and curvature characteristics. In this study, our objective is to explore the influence of a cosmic string on the deflection angle of photon lights and the corresponding effective potential within the framework of a space-time described by the line-element (\ref{1}). We derive the deflection angle for photon light in this particular space-time and carefully analyze the obtained results. Furthermore, we compare our findings with existing results in the literature, allowing for a comprehensive evaluation and a deeper understanding of the impact of cosmic strings on light deflection.

\section{Gravitational lensing in EiBI gravity background with cosmic strings}

In this section, we aim to study the null geodesic phenomena and the deflection angle of light rays within the framework of Eddington-inspired Born-Infeld (EiBI) gravity, while also considering the influence of cosmic strings. Our investigation utilizes the Lagrangian method to derive a one-dimensional energy expression. We aim to analyze the effect of the cosmic string parameter, denoted as $\alpha$, on the effective potential governing both null and time-like geodesics. Finally, we derive the angle of deflection of photon rays and discuss the influences of both the Eddington parameter $\epsilon$ and the cosmic string parameter.

We begin this section with the Lagrangian of a system defined by \cite{aa6,aa7,aa8,aa9,aa25,aa28,aa30,aa31}
\begin{equation}
\mathcal{L}=\frac{1}{2}\,g_{\mu\nu}\,\left(\frac{dx^{\mu}}{d\tau}\right)\,\left(\frac{dx^{\nu}}{d\tau}\right)=\frac{1}{2}\,g_{\mu\nu}\,\dot{x}^{\mu}\,\dot{x}^{\nu}\quad (\mu,\nu=0, \ldots ,3),
\label{3}
\end{equation}
where $\tau$ is the affine parameter of the curve, and $g_{\mu\nu}$ is the metric tensor with $x^{\mu}(=t, r, \phi, z)$. 

Using the line-element (\ref{1}) in $\theta=\frac{\pi}{2}$ hyper-surface, we find
\begin{equation}
\mathcal{L}=\frac{1}{2}\,\Bigg[-\dot{t}^2+\frac{\dot{r}^2}{(1+\frac{\epsilon}{r^2})}+\alpha^2\,r^2\,\dot{\phi}^2\Bigg],
\label{3b}
\end{equation}
where dot represents ordinary derivative w. r. t an affine parameter $\tau$.

There are two constant of motion given by
\begin{eqnarray}
    &&E=\dot{t},\nonumber\\
    &&L=\alpha^2\,r^2\,\dot{\phi}\Rightarrow \dot{\phi}=\frac{L}{\alpha^2\,r^2},
    \label{3c}
\end{eqnarray}
where $E$ is the conserved energy parameter, and $L$ the conserved angular momentum. 

With these, the Lagrangian (\ref{3b}) for light-like or time-like geodesics becomes
\begin{equation}
    \Big(\frac{dr}{d\tau}\Big)^2=\Big(1+\frac{\epsilon}{r^2}\Big)\,\Big(\varepsilon+E^2-\frac{L^2}{\alpha^2\,r^2}\Big),
    \label{3d}
\end{equation}
where $\varepsilon=0$ for null geodesics and $-1$ for time-like geodesics.

One can see from above that the turning point $r=r_0$ in the geometry under consideration occurs for $\frac{dr}{d\tau}=0$, {\it i. e.}, $r_0=\frac{L}{\alpha\,E}$ for null geodesics that depends on the cosmic string parameter. It is better to mention here that for the wormhole case, $\epsilon<0$, the solution has a minimum radius given by $r=r_0=\sqrt{|\epsilon|}$, thus, photons with sufficient energy pass to the other side of the wormhole region. The effective potential expression of the system given by $V_{eff}=\Big(-\varepsilon+\frac{L^2}{\alpha^2\,r^2}\Big)$ depends on the cosmic string parameter. 

In Figure 1 (left panel), we present the effective potential for null geodesics with a fixed value of $L$ and varying $\alpha$, considering two cases: $\alpha =1$ and $\alpha =1/2$. On the right panel of Figure 1, we depict the effective potential for null geodesics while keeping $\alpha$ constant at $1/2$ and varying the values of $L$.

Similarly, in Figure 2 (left panel), we illustrate the effective potential for time-like geodesics with a constant value of $L$, but changing $\alpha$. Here, we examine two scenarios: $\alpha = 1$ and $\alpha =1/2$. Moving to the right panel of Figure 2, we showcase the effective potential for time-like geodesics once more, but this time, we maintain $\alpha$ at a constant value of $1/2$ and explore different values of $L$.

\begin{figure}
    \centering 
    \includegraphics[width=3.0in,height=1.8in]{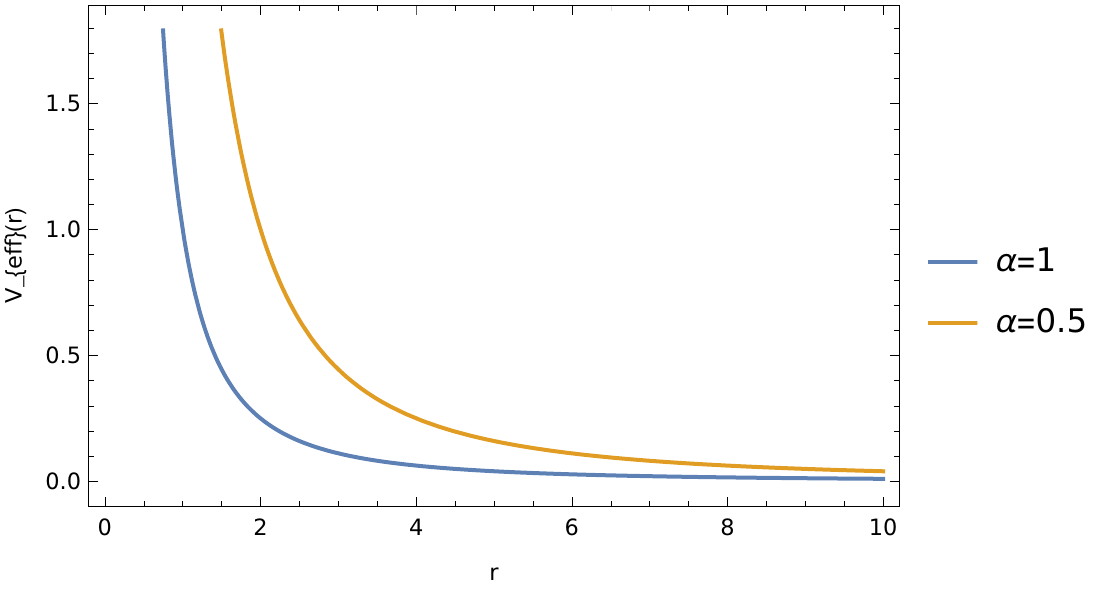}\includegraphics[width=3.0in,height=1.8in]{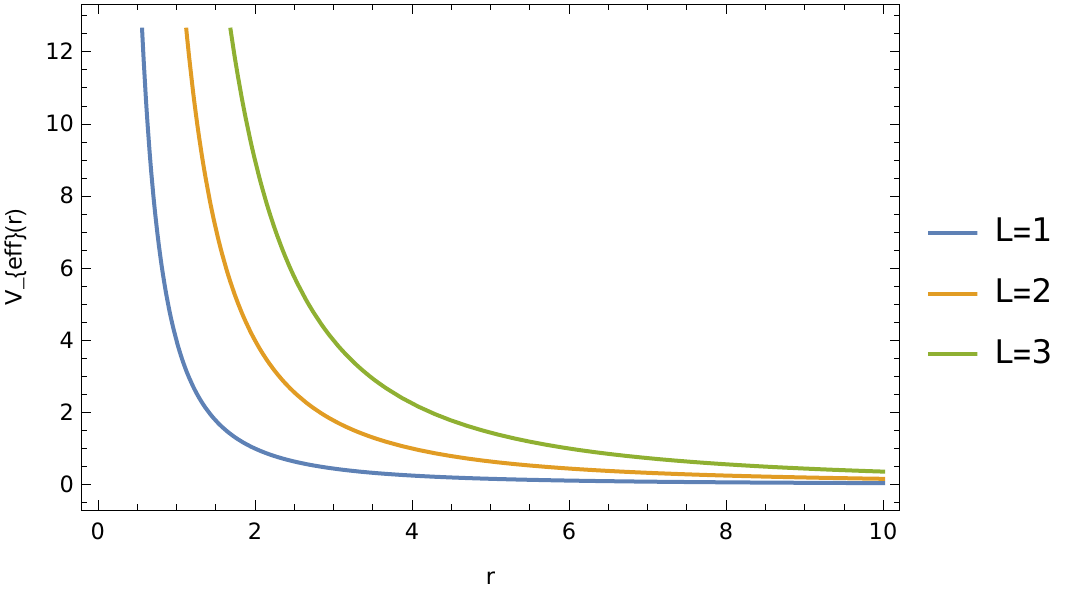}
    \label{fig: 1}
    \caption{Effective potential for null or light-like geodesics. Here $L=1$ at the left figure and $\alpha=1/2$ at the right one.}
\end{figure}

\begin{figure}
    \centering
    \includegraphics[width=3.0in,height=1.8in]{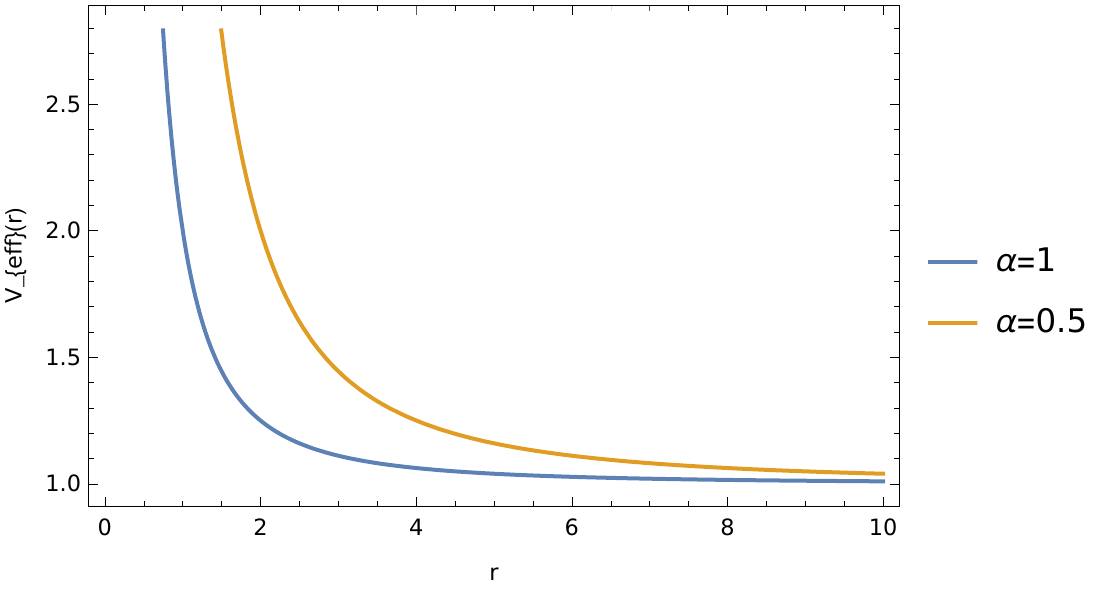}\includegraphics[width=3.0in,height=1.8in]{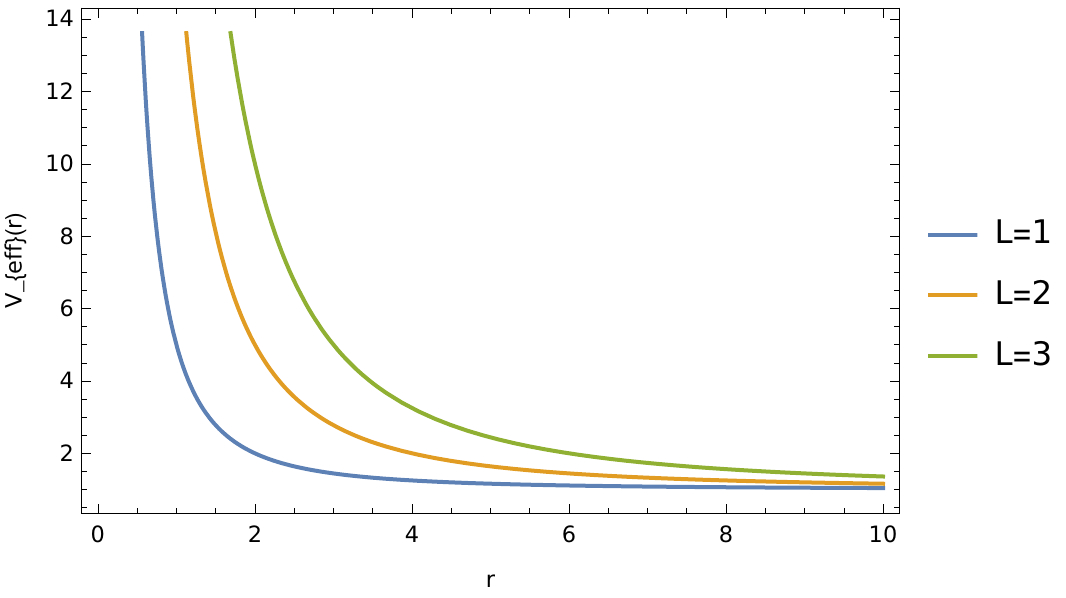}
    \label{fig: 2}
    \caption{Effective potential for time-like geodesics. Here $L=1$ at the left figure and $\alpha=1/2$ at the right one.}
\end{figure}

Now, using Eqs. (\ref{3c}) and (\ref{3d}), we obtain
\begin{equation}
    \frac{d\phi}{dr}=\frac{\dot{\phi}}{\dot{r}}=\frac{\beta}{\alpha^2\,\sqrt{(r^2+\epsilon)\Big(r^2-\frac{\beta^2}{\alpha^2}\Big)}},
    \label{3e}
\end{equation}
where $\beta=\frac{L}{E}$ is the impact parameter characterizing a particular null geodesic with the conserved energy parameter $E$ and the angular momentum $L$ and therefore, $r_0=\frac{\beta}{\alpha}$. By symmetry, the contributions to $\Delta\phi$ before and after the turning point are equal Hence, we can obtain the deflection angle \cite{aa28}
\begin{eqnarray}
\delta\phi=\Delta\phi-\pi,
\label{3p}
\end{eqnarray}
where
\begin{eqnarray}
    \Delta\phi=\frac{2\,\beta}{\alpha^2}\,\int^{\infty}_{r_0}\,\frac{dr}{\sqrt{(r^2+\epsilon)\Big(r^2-\frac{\beta^2}{\alpha^2}\Big)}}=\frac{2\,\beta}{\alpha^2}\,\int^{\infty}_{\frac{\beta}{\alpha}}\,\frac{dr}{\sqrt{(r^2+\epsilon)(r^2-\frac{\beta^2}{\alpha^2})}}.
    \label{3f}
\end{eqnarray}

Defining $z=\frac{\beta}{\alpha\,r}$ and $g=-\frac{\alpha^2\,\epsilon}{\beta^2}$ into the above integral results
\begin{eqnarray}
    \Delta\phi=\frac{2}{\alpha}\,K(g),
    \label{3g}
\end{eqnarray}
where we have defined
\begin{equation}
    K(g)=\int^{1}_{0}\,\frac{dz}{\sqrt{\Big(1-z^2\Big)\Big(1-g\,z^2\Big)}}.
    \label{3h}
\end{equation}
The above expression is valid both for the WH case $\epsilon <0$ and with cosmic string case $\epsilon>0$. However, in topological defect of cosmic string in EiBI gravity background with the positive parameter $\epsilon>0$, we can write $g=-\gamma$, where $\gamma=\frac{\alpha^2\,\epsilon}{\beta^2}>0$. One can easily show that\footnote{
We know that
\begin{equation}
    K(\gamma)=\int^{1}_{0}\,\frac{dz}{\sqrt{\Big(1-z^2\Big)\Big(1-\gamma\,z^2\Big)}}.\nonumber
\end{equation}
Changing the variable $z=\sqrt{1-y^2}$ into the above integral, one can find
\begin{equation}
    K(\gamma)=\int^{1}_{0}\,\frac{dy}{\sqrt{\Big(1-y^2\Big)\Big(1-\gamma+\gamma\,y^2\Big)}}.\nonumber
\end{equation}
Therefore, 
\begin{equation}
    K(-\gamma)=\int^{1}_{0}\,\frac{dy}{\sqrt{\Big(1-y^2\Big)\Big(1+\gamma-\gamma\,y^2\Big)}}=\frac{1}{\sqrt{1+\gamma}}\,\int^{1}_{0}\,\frac{dy}{\sqrt{\Big(1-y^2\Big)\Big(1-\frac{\gamma}{1+\gamma}\,y^2\Big)}}.\nonumber
\end{equation}
Thus, one can write
\begin{equation}
    K(-\gamma)=\frac{1}{\sqrt{1+\gamma}}\,K\Big(\frac{\gamma}{1+\gamma}\Big).\nonumber
\end{equation}
} 
\begin{equation}
    K(g)=K(-\gamma)=\frac{1}{\sqrt{1+\gamma}}\,K\Big(\frac{\gamma}{1+\gamma}\Big).
    \label{3i}
\end{equation}

The deflection angle of photon light therefore is given by
\begin{eqnarray}
\delta\phi=\frac{2}{\alpha}\,K(g)-\pi,
\label{3j}
\end{eqnarray}
where
\begin{eqnarray}
   \delta\phi=\left\{ \begin{array}{rcl}
\frac{2}{\alpha}\,K\Big(-\frac{\alpha^2\,\epsilon}{\beta^2}\Big)-\pi & \mbox{for} & \epsilon <0 \\
\frac{2}{\alpha\,\sqrt{1+\frac{\alpha^2\,\epsilon}{\beta^2}}}\,K\Bigg(\frac{\frac{\alpha^2\,\epsilon}{\beta^2}}{1+\frac{\alpha^2\,\epsilon}{\beta^2}}\Bigg)-\pi & \mbox{for} & \epsilon>0.
\end{array}\right.
\label{3k}
\end{eqnarray}

We can see that the deflection angle of photon light is significantly influenced by the presence of cosmic string parameter $\alpha$. Figure 3 (left panel) displays the deflection angle without cosmic strings, where $\alpha$ is set to 1. In Figure 3 (right panel), we observed the deflection angle for a fixed value of the cosmic string parameter, specifically $\alpha =3/4<1$. It is noteworthy that in the case of a wormhole ($g > 0$), the deflection angle of light diverges as the turning point approaches the throat radius. Conversely, in the case of a wormhole with a cosmic string ($g < 0$), the deflection angle remains finite throughout. Figure 4 presents the deflection angle for various values of the cosmic string parameter, specifically $\alpha = 0.4, 0.6, 0.8$. We have observed that as the value of cosmic string parameter $\alpha$ increases, the deflection angle decreases. This trend highlights the influence of the cosmic string parameter on the deflection behavior of photon light.

\begin{figure}
    \centering
    \includegraphics[width=2.5in,height=1.6in]{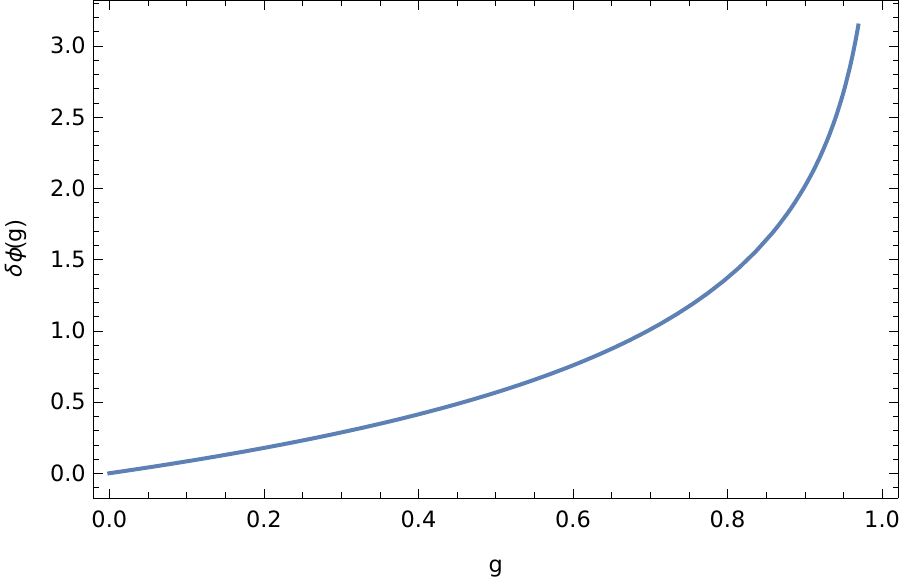}\quad\quad 
    \includegraphics[width=2.4in,height=1.6in]{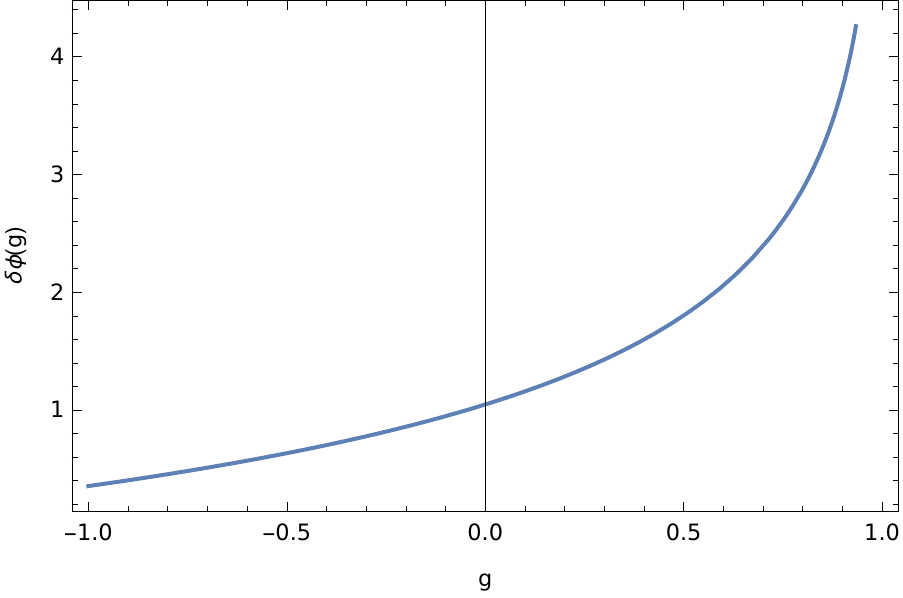}
    \label{fig: 3}
    \caption{Light deflection, $\delta\phi(g)$, as a function of $g$ without and with cosmic string. Here $\alpha=1$ at the left figure and $\alpha=3/4$ at the right one.}
\end{figure}

\begin{figure}
    \centering
    \quad\quad\quad\quad\includegraphics[width=3.0in,height=1.6in]{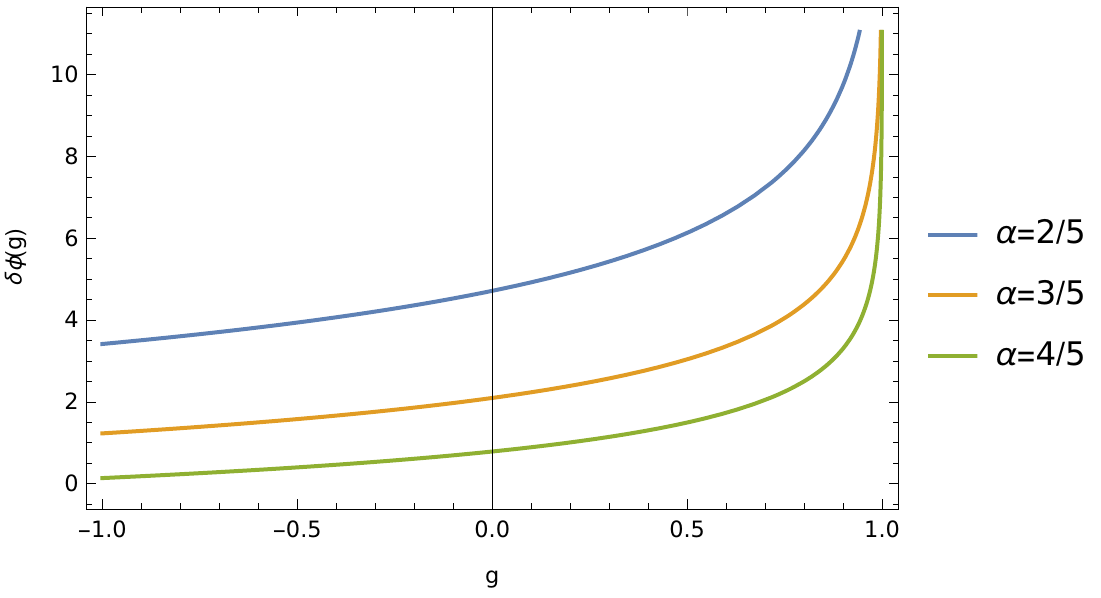}
    \label{fig: 7}
    \caption{Light deflection, $\delta\phi(g)$, as a function of $g$ for different values of $\alpha$.}
\end{figure}

Since the cosmic string parameter lies in the interval $0< \alpha <1$, thus, at the weak field limit, $\beta >\epsilon$, where $g<<1$ or $g \to 0$, one can express the complete elliptic integral $K(g)$ as
\begin{eqnarray}
    K(g)=\frac{\pi}{2}\,\sum^{\infty}_{n=0}\,\Bigg(\frac{(2n!)}{2^{2\,n}\,(n!)^2}\Bigg)^2\,(g)^{n}=\frac{\pi}{2}\,\Big[1-\frac{1}{4}\,\frac{\alpha^2\,\epsilon}{\beta^2}+\frac{9}{64}\,\frac{\alpha^4\,\epsilon^2}{\beta^4}-\mathcal{O} \Big(\frac{\epsilon}{\beta^2}\Big)^3\Big].
    \label{3L}
\end{eqnarray}

Thus, the deflection angle of photon light in this limit becomes
\begin{eqnarray}
    \delta\,\phi&=&\pi\,\Big(\frac{1}{\alpha}-1\Big)-\frac{\pi\,\alpha}{4}\,\Big(\frac{\epsilon}{\beta^2}\Big)+\frac{9\,\pi\,\alpha^3}{64}\,\Big(\frac{\epsilon}{\beta^2}\Big)^2-\mathcal{O}\Big(\frac{\epsilon}{\beta^2}\Big)^3\nonumber\\
    &=&\pi\,\Big(\frac{1}{\alpha}-1\Big)-\frac{\pi\,\alpha}{4}\,\Big(\frac{\epsilon}{\beta^2}\Big)+\mathcal{O}\Big(\frac{\epsilon}{\beta^2}\Big)^2.
    \label{3M}
\end{eqnarray}

For $\epsilon=-b^2<0$, the space-time (\ref{1}) under consideration becomes Morris-Throne-type wormhole with cosmic strings, and the deflection of null geodesics has studied recently in Ref. \cite{aa31}.

For $\alpha \to 1$, one will obtain the following expression from (\ref{3M}) given by 
\begin{equation}
    \delta\,\phi=-\frac{\pi}{4}\,\Big(\frac{\epsilon}{\beta^2}\Big)+\frac{9\,\pi}{64}\,\Big(\frac{\epsilon}{\beta^2}\Big)^2-\mathcal{O}\Big(\frac{\epsilon}{\beta^2}\Big)^3.
    \label{3N}
\end{equation}

The given expression represents the deflection angle for null or light-like geodesics in a wormhole space-time within the framework of Eddington-inspired Born-Infeld (EiBI) gravity background, in the absence of cosmic strings, $\alpha \to 1$. If we substitute the parameter $\epsilon=-b^2$ into the above deflection angle (\ref{3N}), we will get back the same results obtained in an Ellis wormhole background in the weak field limit in Refs. \cite{aa18,aa28}. Furthermore, upon examining Equation (\ref{3M}), we can observe that the first term in the expression corresponds to the deflection caused solely by the cosmic string effect. On the other hand, the second term in the expression represents the combined contribution arising from both the EiBI gravity background and the presence of cosmic strings.

In Ref. \cite{aa33}, the authors investigated the phenomenon of gravitational lensing induced by a topologically charged EiBI monopole or wormhole space-time. They examined this effect in both the weak and strong field limits. In the case of weak field limit, the deflection angle can be expressed as follows (replacing $\alpha \to \alpha_0$ in Eq. (15) in \cite{aa33}):
\begin{eqnarray}
    \delta\phi=\pi\,\Big(\frac{1}{\alpha_0}-1\Big)-\frac{\pi}{4\,\alpha_0}\,\Big(\frac{\epsilon}{\beta^2}\Big)+\mathcal{O}\Big(\frac{\epsilon}{\beta^2}\Big)^2.
    \label{3P}
\end{eqnarray}

Upon examining the expressions for the deflection angle equations (\ref{3M}) and (\ref{3P}), it becomes evident that the angle of deflection experienced by photon rays in the presence of cosmic string effects is greater when compared to the corresponding result (\ref{3P}) obtained under the influence of topological charged space-time in the EiBI gravity background. In essence, the deflection caused by cosmic strings exhibits a more pronounced effect on photon trajectories within the specific framework of EiBI gravity.

\section{Conclusions}

Numerous studies have extensively examined the phenomenon of photon deflection in various curved space-time backgrounds, including those generated by black holes, wormholes, and topological defects. These investigations have involved thorough analyses of the impact of curvature on the angle of deflection experienced by null geodesics.

Our present research focuses on examining the deflection of photon rays within the background of Eddington-inspired Born-Infeld (EiBI) gravity, which incorporates cosmic strings. By employing the Lagrangian method, we have derived a one-dimensional energy expression given by Eq. (\ref{3d}). Our findings demonstrate that the effective potential of the system, applicable to both null and time-like geodesics, is influenced by the cosmic string characterized by the parameter $\alpha$.

We have presented several figures illustrating the effective potential for both null and time-like geodesics within EiBI gravity, considering different values of the cosmic string parameter $\alpha$ and the angular momentum $L$ (see Figs. 1--2). Furthermore, we have analytically derived the deflection angle for photon light or null geodesics, obtaining an expression given by Eq. (\ref{3M}) under the weak field limit. This expression provides insights into the relationship between the deflection angle and relevant parameters in the EiBI gravity scenario.

Upon analyzing expression (\ref{3M}), it becomes evident that the angle of deflection for photon rays is significantly influenced by the cosmic string parameter, denoted as $\alpha$. This parameter introduces a shift in the resulting deflection angle, altering its magnitude compared to the case without the cosmic string, $\alpha \to 1$.

To illustrate the influence of cosmic strings, we have generated Figs. 3--4 showcasing the variation of the deflection angle for different values of the cosmic string parameter $\alpha$. These Figures provide evidence of how the cosmic string parameter impacts the deflection of light rays within EiBI gravity theory.

Moreover, it is noteworthy that the angle of deflection obtained here is greater when compared to the previously established result for a topologically charged EiBI gravity background. This indicates that the deflection of photon light under the influence of the specific conditions (cosmic strings) considered in this study exhibits a more substantial deviation from the known outcomes in the context of topologically charged EiBI gravity.

In forthcoming research endeavors, our attention will be directed towards exploring modified or generalized versions of the EiBI theory and/or Born-Infeld-f(R) theory, as outlined in the studies \cite{SO, SO2}. Our investigation will specifically incorporate the presence of cosmic strings and meticulously analyze the resulting implications and outcomes.

\section*{Data Availability Statement}

No new data were generated or analyzed in this study.

\section*{Conflict of Interest}

No conflict of interests in this study.

\section*{Acknowledgement}

We sincerely acknowledged the referee for his/her valuable comments and helpful suggestions.

\end{document}